\documentclass[pss]{wiley2sp} 
\usepackage{amsmath}
\usepackage{bm}    
\usepackage{amssymb}          
\usepackage{graphicx}
\usepackage{epstopdf}
\usepackage{hyperref}
\begin{document}

\title{Where is the quantum critical point in the cuprate superconductors?}

\titlerunning{Quantum criticality in the cuprate superconductors}

\author{%
  Subir Sachdev\textsuperscript{\Ast}
  }

\authorrunning{S. Sachdev}

\mail{e-mail:
  \textsf{sachdev@physics.harvard.edu}}

\institute{%
  Department of Physics, Harvard University, Cambridge MA 02138, USA
}


\pacs{71.10.Hf, 75.10.Jm, 74.25.Dw, 74.72.-h\\~\\
\href{http://arxiv.org/abs/0907.0008}{arXiv:0907.0008}\\~\\
{\bf Talk at the Conference on Quantum Criticality and Novel Phases, Dresden,  August 2-5,  2009}\\~\\
Slides at \href{http://qpt.physics.harvard.edu/talks/qcnp09.pdf}{http://qpt.physics.harvard.edu/talks/qcnp09.pdf} } 

\abstract{%
%
%
%
\abstcol{%
Transport measurements in the hole-doped cuprates show a `strange metal'
normal state with an electrical resistance which varies linearly with temperature.
This strange metal phase is often identified with the quantum critical region
of a zero temperature quantum critical point (QCP) at hole density $x=x_m$, near optimal doping.
A long-standing problem with this picture is that  
low temperature experiments within the superconducting phase have not shown convincing
signatures of such a optimal doping QCP (except in some cuprates with small superconducting critical temperatures). 
I review theoretical work which proposes a simple
resolution of this enigma.}{ The crossovers in the normal state are argued to be controlled by
a QCP at $x_m$
linked to the onset of spin density wave (SDW) order in a ``large'' Fermi surface metal, leading
to small Fermi pockets for $x<x_m$. A key effect is that the onset of superconductivity at low temperatures
disrupts the simplest canonical quantum critical crossover phase diagram. In particular, the competition
between superconductivity and SDW order {\em shifts} the actual QCP to a lower doping $x_s < x_m$
in the underdoped regime,
so that SDW order is only present for $x<x_s$. I review the phase transitions and crossovers
associated with the QCPs at $x_m$ and $x_s$: the resulting phase diagram as a function of $x$, temperature, and applied
magnetic field consistently explains a number of recent experiments.}}

%
%

\maketitle   

\section{Introduction}

\begin{figure}[tb]%
\includegraphics*[width=0.5\textwidth]{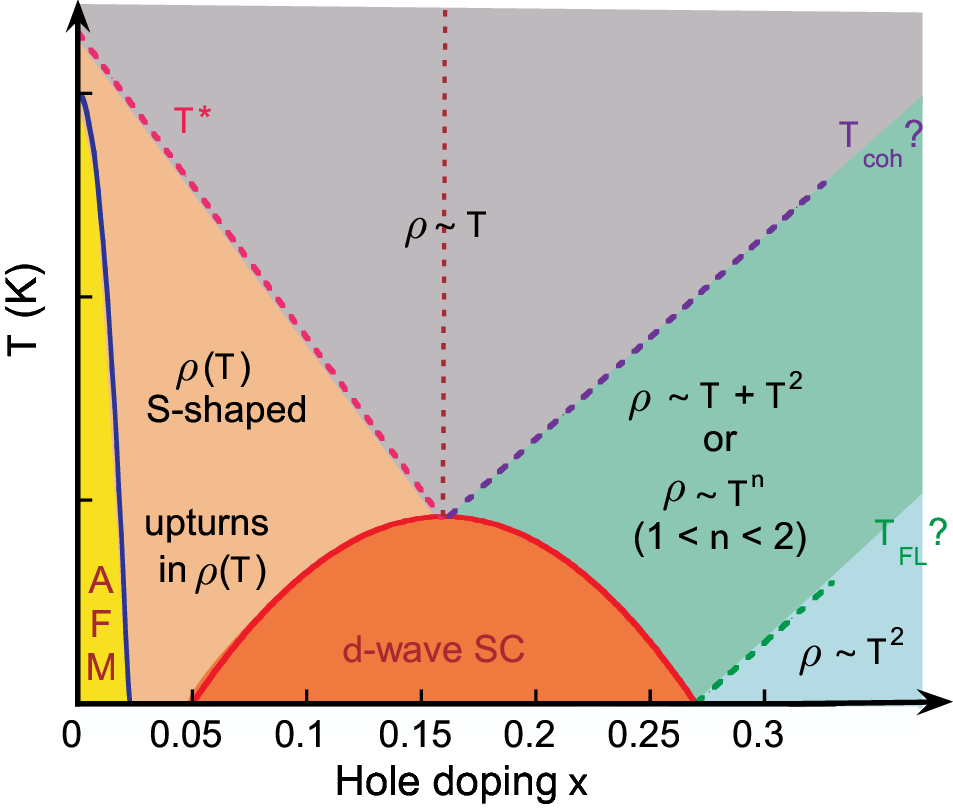}
\caption{From Ref.~\cite{hussey}: Crossover phase diagram for the resistivity ($\rho$) in the hole doped cuprates. The strange metal phase is the regime above optimal doping where $\rho \sim T$.}
\label{fighussey}
\end{figure}

The strange metal phase of the hole-doped cuprates is often regarded as
the central mystery in the theory of high temperature superconductivity. This appears
near optimal doping, and exhibits a resistivity which is linear in temperature ($T$)
over a wide temperature range. A useful review of the transport data has been
provided recently by Hussey \cite{hussey}: we show his crossover phase diagram in Fig.~\ref{fighussey}.

It is tempting to associate the strange metal with the finite temperature quantum critical
region linked to a $T=0$ quantum critical point (QCP) \cite{sy}: this is the regime where $k_B T$ is the most important perturbation away from the QCP. From the sketch in Fig.~\ref{fighussey}, we see that
such a QCP should be at a hole doping $x=x_m$, near optimal doping \cite{varma}. However, experiments
have not so far revealed any clear-cut quantum phase transition at low $T$ in the superconducting state of cuprates
near such values
of the hole doping $x$.
(A notable exception is the work of Panagopoulos and collaborators \cite{christos1,christos2}, which I will discuss shortly. Also, the situation in materials with a small $T_c$ is simpler, and recent experiments \cite{nlsco1} on Nd-LSCO do show
an optimal doping QCP; these experiments provide support for the ideas presented below, as will
be discussed at the end of Section~\ref{sec:phase}.)

One often-stated resolution of this puzzle is that the order parameter associated with optimal-doping
QCP is difficult to detect \cite{varma,sudip}. This could be because it involves subtle forms of symmetry breaking,
or it is associated with a `topological' transition which cannot be characterized by a local order parameter.

Here, I review an alternate and simple resolution which has appeared in recent theoretical work \cite{rkk3,gs,moon} based 
upon a QCP associated with the onset of the clearly observed spin density wave (SDW) order; this work
builds on a number of important new experimental observations \cite{nlsco1,doiron,cooper,nigel,cyril,suchitra,louis,suchitra2,mesot,mesot2,mesot3,nlsco2,nlsco3}.
In this theory, there is no $T=0$ SDW QCP in the superconducting state at optimal doping, explaining its experimental 
ephemerality. The normal state crossovers in Fig.~\ref{fighussey}
are controlled by an optimal doping SDW QCP at $x=x_m$ which can be directly observed when the system remains a metal at $T=0$ 
in the presence of a strong enough magnetic field 
(this is the reason for the subscript $m$); this proposal was also made in Ref.~\cite{nlsco1}. As $T$ is lowered towards
this metallic SDW QCP, there is an onset of $d$-wave superconductivity, One of the key consequences
of superconductivity is that it {\em shifts the SDW QCP towards the underdoped regime}, to a hole
density $x=x_s < x_m$. So, in the absence of an applied magnetic field, the directly observable SDW 
QCP is at $x=x_s$, and it only controls the crossover within the superconducting state. 

Panagopoulos and collaborators \cite{christos1,christos2} have used 
 muon spin relaxation and ac-susceptibility measurements on a series of pure and Zn-substituted hole-doped cuprates
 to observe a glassy slowing down of spin fluctuations below optimal doping, and have argued that these results
 provide experimental evidence for a quantum transition. We believe their observations reflect the underlying SDW QCP in 
 the metal at $x=x_m$.

Our discussion below is oriented towards the hole-doped cuprates. However, our ideas also apply to the electron-doped
cuprates; indeed, for this case there is significant evidence that $x_s < x_m$, as we will describe in Section~\ref{sec:edoped}.

There has also been discussion in the literature of optimal doping QCPs associated with experimental
signatures \cite{trsb1,trsb2,trsb3} of 
time-reversal symmetry breaking. We will not comment on these issues here, apart from suggesting
that these may be related to ancillary instabilities to the primary phenomena discussed below.

\section{Phase diagrams}
\label{sec:phase}

\begin{figure}[b]%
\includegraphics*[width=0.5\textwidth]{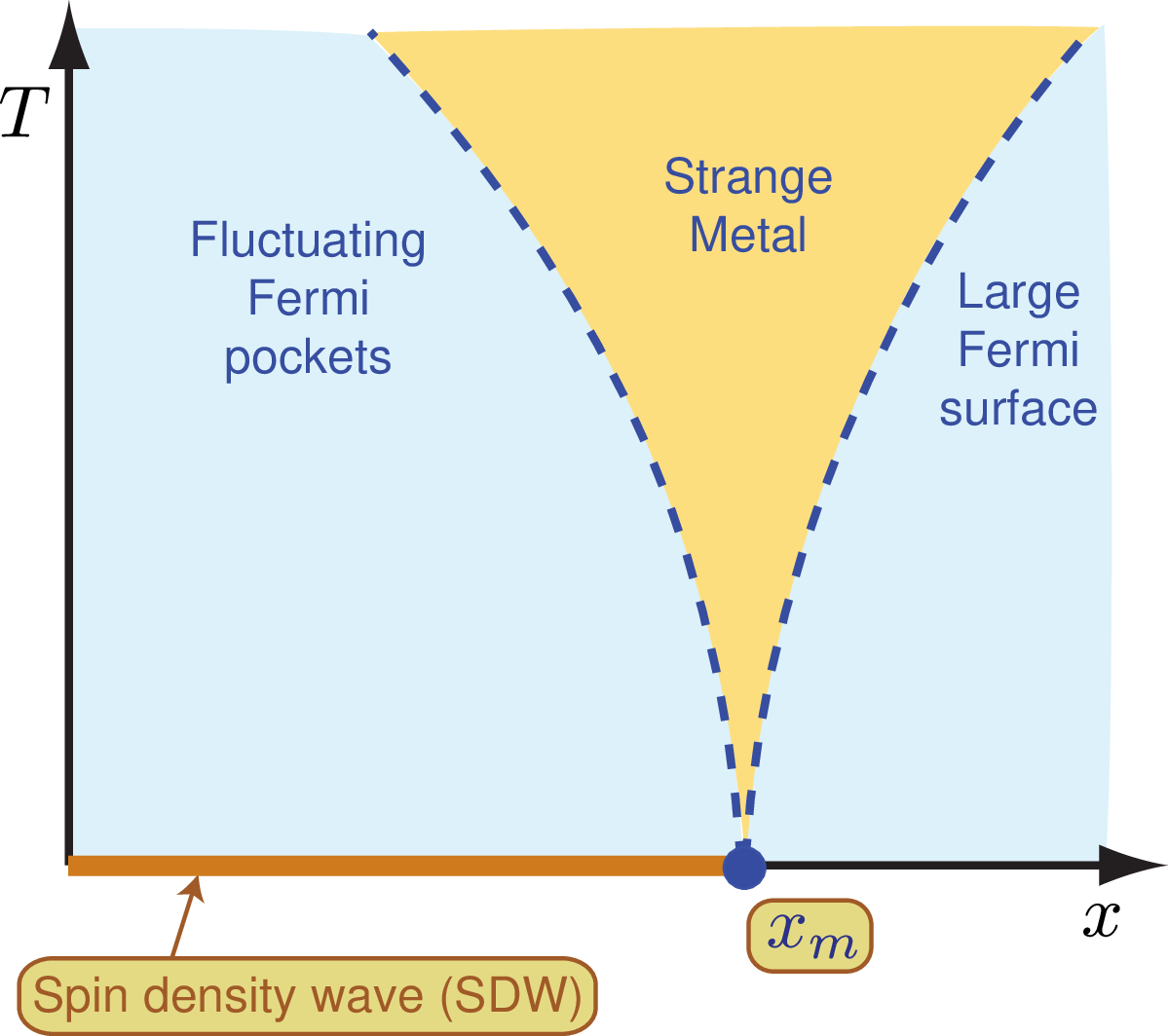}
\caption{The SDW QCP in the metal and its crossover regimes.}
\label{figcross1}
\end{figure}

We begin the review by considering the SDW QCP in the metal, and its associated crossovers; these
are shown in Fig.~\ref{figcross1}.
For $x>x_m$, the ground state is a Fermi liquid with a $T^2$ resistivity and a large Fermi surface;
{\em i.e.} the Fermi surface is that obtained from the underlying band structure with no broken symmetries, and it is a large hole-like surface centered at $(\pi, \pi)$ which encloses
an area equal to $2 \pi^2 (1 + x)$. For $x<x_m$, there is the onset of SDW order. As we noted above,
the true ground state is a superconductor, and so we have not yet given a prescription to determine
the precise value of $x_m$: we defer this to the discussion below. Because the SDW order breaks
the non-Abelian spin rotation symmetry, true long-range order can only be present at $T=0$
in two spatial dimensions. The large Fermi surface is broken apart by the SDW order into
electron and hole Fermi pockets, all with a `small' area of order $x$ \cite{sokol}. Remnants of this 
small Fermi pocket structure will survive at $T>0$ and $x<x_m$, as we will discuss below.
In between the small and large Fermi surface regimes, is the quantum-critical strange metal,
as depicted in Fig.~\ref{figcross1}. We will not review here the nature of the QCP at $x=x_m$,
and its associated quantum criticality: there has been a great deal of theoretical work and debate
on this \cite{vrmp}, and under suitable conditions a resistivity linear in $T$ can be obtained. It is 
possible that other order parameters linked to the SDW order are important for transport in the strange metal
regime: in particular, an ``Ising nematic'' order \cite{nematic} couples efficiently to fermions at
all points on the Fermi surface. Other $T=0$ metallic phases with topological order and violations
of the Luttinger theorem (`algebraic charge liquids') \cite{rkk3,rkk2} may also appear as intermediate phases in Fig,~\ref{figcross1},
and are not shown.

\begin{figure}[htb]%
\includegraphics*[width=0.5\textwidth]{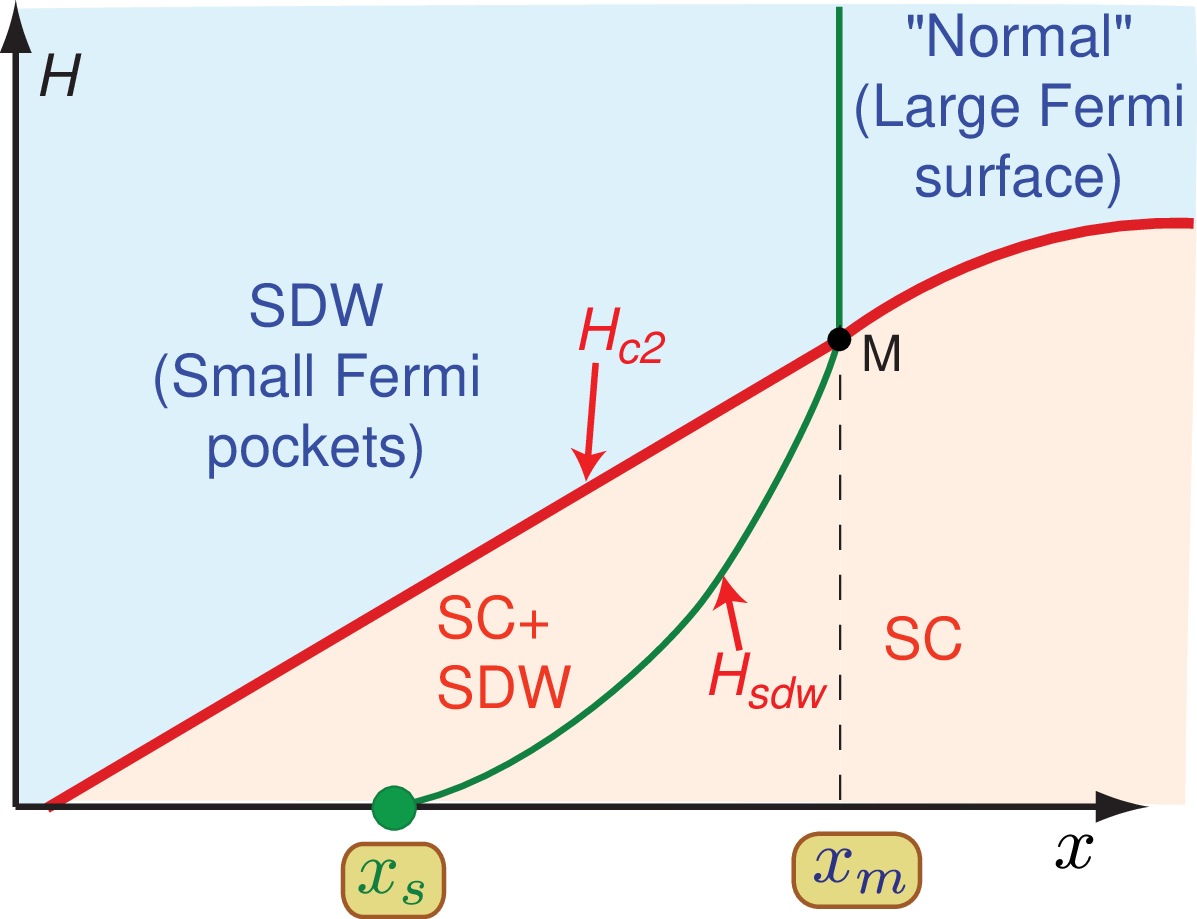}
\caption{From Ref.~\cite{demler}: $T=0$ phase diagram as a function $x$ and $H$ describing
the competition between the SDW and superconducting (SC) orders. There is no superconductivity for $H> H_{c2}$. The position of the field-induced
onset of SDW order within the superconducting state is denoted by $H_{sdw}$. The dashed line only serves to identify $x_m$, and does not represent any transition or crossover.}
\label{figdemler}
\end{figure}

We now turn to the consequences of the onset of superconductivity. It is useful to do this first
using the phenomenological approach in the early work of Ref.~\cite{demler}. We consider the
phase diagram at $T=0$ as a function of $x$ and a magnetic field, $H$, applied perpendicular
to the CuO$_2$ layers. This is obtained using a Landau-Ginzburg action functional expressed in
terms of the SDW and superconducting order parameters; there is a repulsive term between the 
modulus squared of these orders, which represents a `competition' between them. The phase diagram
of Ref.~\cite{demler} is shown in Fig.~\ref{figdemler}.
At small $H$, we have superonductivity, which can co-exist with SDW order at small $x$.
At large $H$, we obtain the non-superconducting states, both with and without SDW order.
The Landau-Ginzburg approach of Ref.~\cite{demler} cannot specify the nature of
the charged excitations in these large $H$ states. Here, as was also done in Refs.~\cite{fuchun,millisnorman,harrison}, we identify
these with the metallic states of Fig.~\ref{figcross1}, as is indicated within parentheses in Fig.~\ref{figdemler}. This leads us to specify the value of $x_m$ by the position of the multicritical point M in Fig.~\ref{figdemler}. Naturally, $x_s$ is the position of the onset of SDW order at $H=0$, $T=0$,
as indicated in Fig.~\ref{figdemler}. A central feature of the phase diagram of Ref.~\cite{demler} is that
$x_s < x_m$, and this
incorporates the physics of competition between SDW order and superconductivity.
One of the consequences of the separation between $x_s$ and $x_m$ is the existence of the
line of quantum phase transitions represented by $H_{sdw}$. This is the locus of the onset of 
field-induced SDW order within the superconducting state. The existence of the critical field $H_{sdw}$ was
predicted in Ref.~\cite{demler}, and has since been observed in neutron scattering and muon
spin resonance experiments in LSCO \cite{boris,mesot,mesot2}, and more recently in YBCO \cite{mesot3}. Furthermore, we can place the recent quantum oscillation experiments \cite{doiron,cooper,nigel,cyril,suchitra,louis}, 
which display convincing evidence of small Fermi pockets, within the SDW phase of Fig.~\ref{figdemler};
specific evidence that the pockets are due to SDW ordering has appeared recently \cite{suchitra2}.

\begin{figure}[hbt]%
\includegraphics*[width=0.5\textwidth]{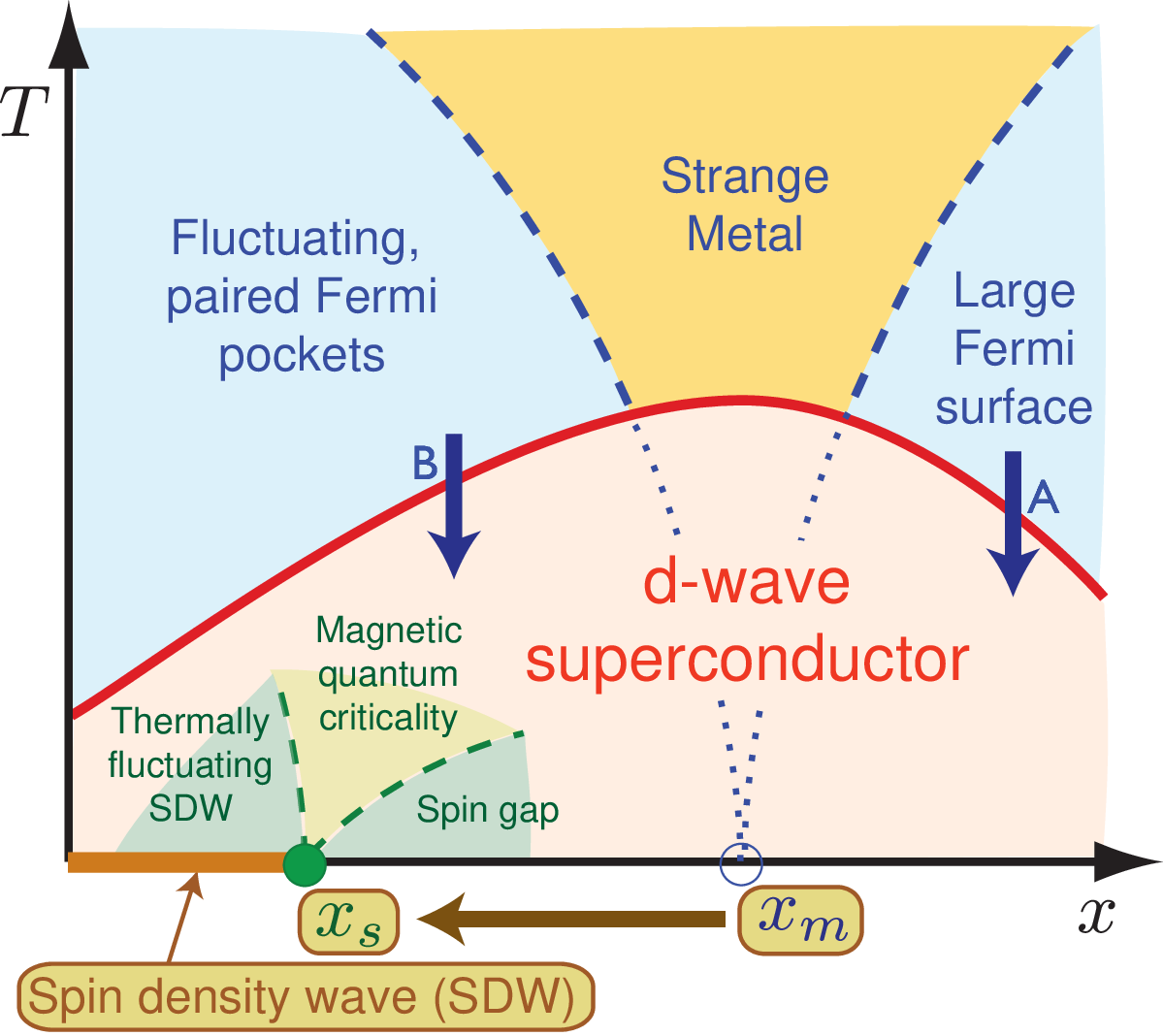}
\caption{Proposed phase diagram for the hole-doped cuprates at $H=0$ from Ref.~\cite{moon}. There is no QCP at $x_m$
(the dotted lines do not denote any crossovers),
and its location only identifies the hole density of point M in Fig.~\ref{figdemler}.
The QCP for the onset of SDW order is at $x_s$, also identified in Fig.~\ref{figdemler}, 
and it controls crossovers within the superconducting state.}
\label{figcross}
\end{figure}

\begin{figure}[b]%
\includegraphics*[width=0.5\textwidth]{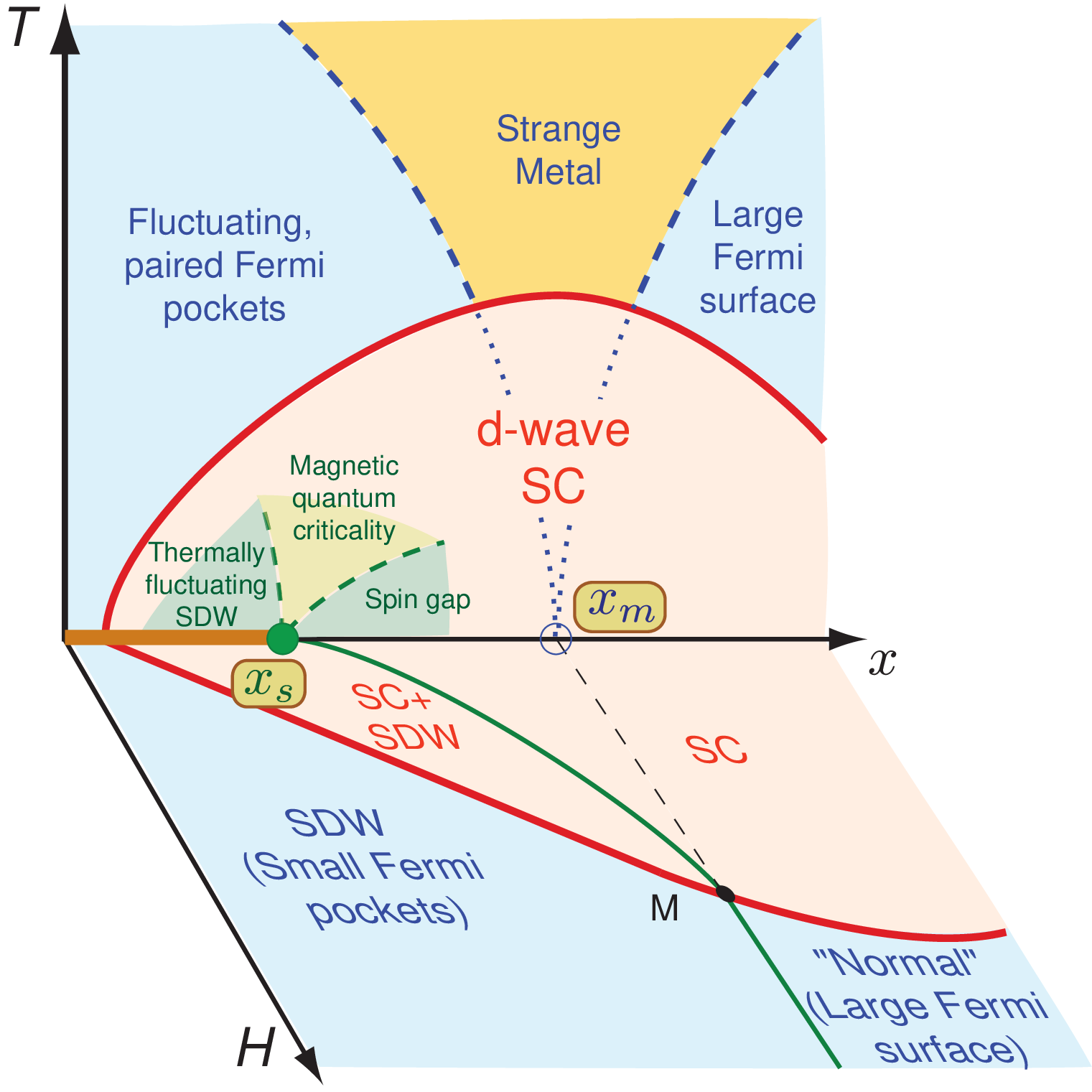}
\caption{The combination of Figs.~\ref{figdemler} and~\ref{figcross}.}
\label{figmerge}
\end{figure}

Finally, we combine the physics of Figs.~\ref{figcross1} and~\ref{figdemler} to describe
the interplay between superconductivity and SDW order at $T \geq 0$, but $H=0$.
The proposal of Ref.~\cite{moon} appears in Fig.~\ref{figcross}.
We also show combination of Figs.~\ref{figdemler} and~\ref{figcross} in the unified phase diagram in Fig.~\ref{figmerge}.
Above the superconducting $T_c$, the crossovers are controlled by the metallic QCP at $x=x_m$
in Figs.~\ref{figcross1} and~\ref{figdemler}. However, when we move below $T_c$, the onset of superconductivity
moves the actual QCP to $x=x_s$, and so the attention shifts to the QCP in the underdoped regime
within the superconducting phase.

A notable feature of Fig.~\ref{figcross} is the complex nature of the crossovers as temperature
is lowered in the underdoped regime, {\em e.g.\/} along the arrow labeled B. For $T>T_c$, the metallic state
acts as if it is on the ordered side of a SDW transition. Thus, the Fermi pockets are locally formed,
but are disrupted only by thermal fluctuations of the SDW order in the ``renormalized classical'' regime.
There is a similarity here to the physics of the ``fluctuating stripe'' approach \cite{stripermp}. 
However, upon moving to $T<T_c$, the superconducting state is on the ``quantum disordered''
side of a SDW ordering transition. Thus the collective SDW excitations eventually acquire a spin gap at
the SDW ordering wavevector \cite{spingap}. 
Such a crossover is especially relevant in understanding experiments
in the YBCO class of superconductors, which have not displayed long-range SDW order
in the superconducting phase, but instead have a spin gap. Nevertheless, as shown in Fig.~\ref{figcross}, we can regard
the underdoped metallic state above 
$T_c$ as having thermally fluctuating SDW order. Similarly, as is clear
from Fig.~\ref{figdemler}, YBCO can have a SDW ordered state with small Fermi pockets
at high magnetic fields, as is needed to understand the quantum oscillation experiments  \cite{doiron,cooper,nigel,cyril,suchitra,louis,suchitra2}.

It is quite likely that there is valence bond solid (VBS) and/or 
Ising nematic order in the superconducting state for $x_s < x < x_m$ (see Fig.~\ref{figtsdw} below). 
In particular, mechanisms for
valence-bond stripe order have been discussed in previous work \cite{rkk3,mvreview}: after the onset of
superconductivity, the algebraic charge liquid
phases near $x_m$ (noted earlier) are unstable to a confinement transition
which can lead to breaking of the square lattice space group symmetry. If so, the left crossover line emerging 
from $x_m$ in Fig.~\ref{figcross} would become an actual
thermal phase transition at which lattice symmetry is restored.

We also note the different situation in a low $T_c$ superconductor, such as Nd-LSCO \cite{nlsco1,nlsco2,nlsco3}.
Here, the superconductivity is not well formed, and so there is no significant shift from $x_m$ to $x_s$. Further, by
applying a moderate magnetic field, it is possible to suppress superconductivity and 
investigate the quantum phase transition between the metallic SDW and ``Normal'' states of Fig.~\ref{figdemler}.
Such experiments \cite{nlsco1,nlsco2,nlsco3} reveal an optimal doping $x_m$, with signatures in transport properties
of the transition from the ``small Fermi pockets'' to the ``large Fermi surface'' states of Fig.~\ref{figcross1},
thus providing significant support for the present theory.

\section{Electron-doped cuprates}
\label{sec:edoped}

We briefly note here the experimental situation in the electron-doped superconductors.

Recent magnetic quantum oscillation experiments in the field-induced normal state of Nd$_{2-x}$Ce$_x$CuO$_4$ 
by Helm {\em et al.} \cite{helm}
find a sudden change in the oscillation frequency at between $x=0.16$ and $x=0.17$, corresponding to a
transition from small Fermi pockets to a large Fermi surface; this identifies $x_m \approx 0.165$. Consistent with this,
earlier transport measurements by Dagan {\em et al.} \cite{dagan} in the closely related compound Pr$_{2-x}$Ce$_x$CuO$_4$
were argued to indicate a SDW ordering transition for $H> H_{c2}$ at $x_m = 0.165$.
In contrast, neutron
scattering measurements by Motoyama {\em et al.} \cite{motoyama} in the zero field superconducting state of Nd$_{2-x}$Ce$_x$CuO$_4$ 
find that long-range SDW order is present only for $x < x_s \approx 0.13$. Thus $x_s < x_m$, just as required by our theory.

\section{Electronic theory}

Our phase diagrams have so far been obtained by general arguments and phenomenological computations.
Now we show that they can also be realized in a simple electronic model. 

We consider the onset of superconductivity along the arrows labeled A and B  in Fig.~\ref{figcross}, in turn. 

\subsection{Overdoped}

Along arrow A, superconductivity appears from a large Fermi surface state. The SDW fluctuations mediate a 
$d$-wave pairing
interaction along this Fermi surface, as has been discussed in detail in Refs.~\cite{dwave,acs}.
We do not have anything further to say about the resulting $d$-wave superconductivity, because the
earlier theory applies directly in this portion of our phase diagram.

\subsection{Underdoped}

Along arrow B, superconductivity appears from a fluctuating Fermi pocket state. Here, we believe a different
approach is necessary: it is important to incorporate the pocket structure of the single particle dispersion,
before considering the pairing instability to superconductivity. A detailed theory for doing this was
presented in Refs.~\cite{rkk2,rkk3,gs,moon}, and here we outline the simplest ingredients. We will use a weak-coupling
perspective here for simplicity \cite{moon}, although the same theory can also be obtained from a strong-coupling theory
departing from a doped Mott insulator \cite{rkk2,rkk3,gs}.

We focus on the portion of Brillouin zone where $d$-wave pairing amplitude is the largest:
the points $(\pi, 0)$ and $(0, \pi)$.
Let us write $c_{(0,\pi)\alpha} = c_{1\alpha}$, $c_{(\pi,0)\alpha} = c_{2 \alpha}$ where $\alpha = \uparrow,\downarrow$
is the spin index. Also, we denote the single particle energy by
$\varepsilon_{(0,\pi)} = \varepsilon_{(\pi,0)} = \varepsilon_0$. We now want to couple these states to the fluctuating
SDW order. First, we take uniform SDW state polarized in the $z$ direction. Thus for the SDW order
parameter, $\vec{\varphi}$, we have $\vec{\varphi} = (0,0,\varphi)$ with $\varphi>0$. Now the Hamiltonian for the single
particle states coupled to static SDW order is
\begin{eqnarray}
&& H_0 + H_{\rm sdw} = \varepsilon_0 \left( c^\dagger_{1\alpha} c_{1 \alpha} + c_{2 \alpha}^\dagger
c_{2 \alpha} \right) \nonumber \\ &&~~~~~+ \varphi  \left( c^\dagger_{1 \uparrow} c_{2 \uparrow} - c_{1 \downarrow}^\dagger c_{2 \downarrow} + c^\dagger_{2 \uparrow} c_{1 \uparrow} - c_{2 \downarrow}^\dagger c_{1 \downarrow} \right) 
\end{eqnarray}
We diagonalize this by writing
\begin{eqnarray}
H_0 + H_{\rm sdw} &=& (\varepsilon_0 - \varphi)\left( g_+^\dagger g_+ + g_-^\dagger g_- \right) \nonumber \\
&+& (\varepsilon_0 + \varphi)\left( h_+^\dagger h_+ + h_-^\dagger h_- \right)
\end{eqnarray}
where
\begin{eqnarray} 
c_{1 \uparrow} &=& ( g_+ + h_+)/\sqrt{2} \nonumber \\
c_{2 \uparrow} &=& ( g_+ - h_+)/\sqrt{2} \nonumber \\
c_{1 \downarrow} &=& ( g_- + h_-)/\sqrt{2} \nonumber \\
c_{2 \downarrow} &=& ( -g_- + h_-)/\sqrt{2} \label{zg}
\end{eqnarray}
Our main approximation below will be to ignore the high energy fermions $h_\pm$, 
although it is not difficult to extend our formalism to include them, as will be described in forthcoming work. 
The neglect of the $h_\pm$ states
means that we have locally projected our electronic states into the Fermi pocket states
of the SDW order.

Let us now extend the ansatz in Eq.~(\ref{zg}) to an arbitrary spacetime
dependent orientation of the SDW order parameter. To do this, we 
write the SDW order parameter in terms of a bosonic spinor field $z_\alpha$ with
\begin{equation}
\vec{\varphi} = z_\alpha^\ast \vec{\sigma}_{\alpha\beta} z_\beta ,
\end{equation}
where $\vec{\sigma}$ are the Pauli matrices. We can now use the $z_\alpha$
to perform a spacetime-dependent SU(2) rotation on Eq.~(\ref{zg}), and after
dropping the $h_\pm$ and the unimportant factor of $\sqrt{2}$, we obtain our final ansatz
\begin{eqnarray}
c_{1 \uparrow} &=& z_\uparrow g_+ - z_\downarrow^\ast g_- \nonumber \\
c_{2 \uparrow} &=&   z_\uparrow g_+ + z_\downarrow^\ast g_- \nonumber \\
c_{1 \downarrow} &=&  z_\downarrow g_+ + z_\uparrow^\ast g_- \nonumber \\
c_{2 \downarrow} &=& z_\downarrow g_+ - z_\uparrow^\ast g_- . \label{czg}
\end{eqnarray}
Our final theory will be expressed in terms of the spinless fermions $g_\pm$
and the bosonic spinor $z_\alpha$.
Note that Eq.~(\ref{czg}) is invariant under the U(1) gauge transformation 
\begin{equation}
z_\alpha \rightarrow e^{i \phi} z_\alpha~~~;~~~g_+ \rightarrow e^{-i \phi} g_+~~~;~~~g_- \rightarrow e^{i \phi} g_-  , 
\end{equation}
and so this invariance must be obeyed by the effective action for $z_\alpha$ and $g_{\pm}$.

As discussed in some detail in Ref.~\cite{moon}, in the resulting theory, the $g_\pm$ are unstable to a
simple $s$-wave pairing with
\begin{equation}
\langle g_+ g_- \rangle = \Delta .
\end{equation}
For the electron operators, we can use Eq.~(\ref{czg}) to deduce that this pairing implies
\begin{eqnarray}
\left\langle c_{1 \uparrow} c_{1 \downarrow} \right \rangle &=& \Delta \left\langle |z_\alpha|^2 \right \rangle \nonumber \\
\left\langle c_{2 \uparrow} c_{2 \downarrow} \right \rangle &=& - \Delta \left\langle |z_\alpha|^2 \right \rangle; \label{gauge}
\end{eqnarray}
in other words, the physically measurable pairing amplitude has the needed $d$-wave signature.

Finally, we are ready to present the Lagrangian for our minimal universal theory for competition between
superconductivity and SDW order \cite{rkk3,gs,moon}
\begin{eqnarray}
\mathcal{L} &=& \mathcal{L}_z + \mathcal{L}_g \nonumber \\
\mathcal{L}_z &=& \frac{1}{t} \Bigl[ |( \partial_\tau - i A_\tau ) z_\alpha
|^2 + v^2 |({\nabla} - i {\bf A}) z_\alpha |^2 \nonumber \\
&~&~~~~~~~~+ i \lambda ( |z_\alpha|^2 - 1) \Bigr] \label{l} \\
\mathcal{L}_g &=& g_+^\dagger \left[ (\partial_\tau - i A_\tau)  - \frac{1}{2m^*} ( { \nabla} - i {\bf A}  )^2 - \mu  \right] g_+
\nonumber \\ &~&~~+ g_-^\dagger \left[ (\partial_\tau + i A_\tau)  - \frac{1}{2m^*} ({  \nabla} + i {\bf A} )^2 - \mu \right] g_-  .\nonumber
\end{eqnarray}
Apart from the $g_\pm$ and the $z_\alpha$, this theory has two auxilliary fields: \\
({\em i\/}) A Lagrange multiplier $\lambda$ which enforces a unit length constraint on the $z_\alpha$; this
ensures that the magnitude of $\vec{\varphi}$ is fixed, and we are describing orientational
fluctuations of the SDW order.\\
({\em ii\/}) An emergent U(1) gauge field $(A_\tau, {\bf A})$, which is a consequence of 
the invariance in Eq.~(\ref{gauge}); this will mediate the primary interaction between the $z_\alpha$
and the $g_\pm$.

We now use the theory in Eq.~(\ref{l}) to derive the main ingredient used to construct the phase diagrams
of Section~\ref{sec:phase}: the shift in the onset of SDW order due to superconductivity, leading to $x_s < x_m$.
We will assume that the coupling $t$, which determines the strength of the SDW fluctuations, is a monotonically
increasing function of $x$, and so will establish that the critical value $t=t_c$ at the SDW ordering transition obeys
$t_c (\mbox{superconductor}) < t_c (\mbox{metal})$. The value of $t_c$ was computed in Refs.~\cite{ribhu,moon}
in a $1/N$ expansion, in a model where $z_\alpha$ had $N$ spin components. To order $1/N$, the result is \cite{moon}
\begin{eqnarray}
&& \frac{1}{t_c } = \frac{1}{t_c^0} + \frac{1}{N}  \int \frac{d^2 q d \omega}{8 \pi^3} \frac{q^2}{8 (\omega^2 + v^2 q^2)^{1/2}}
\label{tc} \\
&& \times \Biggl[
\frac{1}{(\omega^2 + v^2 q^2) D_1 (q, \omega) } +
\frac{1}{q^2 D_2 (q, \omega) + \omega^2 D_1 (q, \omega)} \Biggr]. \nonumber
\end{eqnarray}
The leading value, $t_c^0$, is insensitive to the presence of the $g_\pm$ fermions, and is a property of $\mathcal{L}_z$ alone.
At order $1/N$, we have omitted a contribution from the $\lambda$ fluctuations which is also insensitve to the fermions,
and displayed only the gauge fluctuation contribution, 
where $D_1$ and $D_2$ are the longitudinal and transverse gauge propagators.
Specifically, the $(A_\tau, {\bf A})$ fluctuations are controlled by an effective action of the form
\begin{eqnarray}
S_A &=& \frac{N}{2}  \int \frac{d^2 q d \omega}{8 \pi^3}
\left[ \left( q_i A_\tau - \omega A_i \right)^2 \frac{D_{1} (q,
\omega)}{q^2}  \right. \nonumber \\
&~&~~~~~~~ \left. + A_i A_j \left( \delta_{ij} - \frac{q_i q_j}{q^2}
\right) D_{2} (q, \omega) \right], \label{sa}
\end{eqnarray}
with $i,j=x,y$, and the values of $D_{1,2}$ are determined by polarization contributions from both $z_\alpha$
and $g_\pm$. In the metal, the gauge fluctuations are screened by the Fermi surfaces of the $g_\pm$ fermions.
In the superconductor,
the opening of the fermion gap, $\Delta$, decreases screening of gauge fluctuations, and this is realized by a 
decrease in the values of $D_{1,2}$ at low momenta and frequencies.
Consequently, gauge fluctuations are {\em enhanced\/} in the superconductor, and 
we see from Eq.~(\ref{tc}) that $t_c$ is a monotonically decreasing function of $\Delta$.
So we see that there is a suppression
of SDW order as $\Delta$ increases, realizing the competition between superconductivity and SDW order.
We have thus established the needed result, that $t_c (\mbox{superconductor}) < t_c (\mbox{metal})$.

Let us restate the above result in more physical terms. The role of the gauge field is to impose the local constraint
associated with the proximity to the Mott insulator. With the onset of superconductivity, the local electronic spin is 
partly absorbed into the singlet Cooper pairs of the superconductor. The enhanced gauge field fluctuations
signify the reduction in the electronic moment available for magnetic order, thus suppressing the SDW ordering transition.

\section{Addendum}
\label{sec:remarks}

Discussions at the QCNP conference raised issues which I address in this addendum.
\begin{itemize}
\item Suchitra Sebastian and Gil Lonzarich pointed out their observation \cite{gil} of a metal-insulator transition with decreasing doping
within the high-field SDW phase of Fig.~\ref{figdemler}. This requires another transition line within this phase in our phase diagrams,
representing the localization transition of the small Fermi pockets, as we have sketched in Fig.~\ref{figloc}.
\begin{figure}%
\includegraphics*[width=0.5\textwidth]{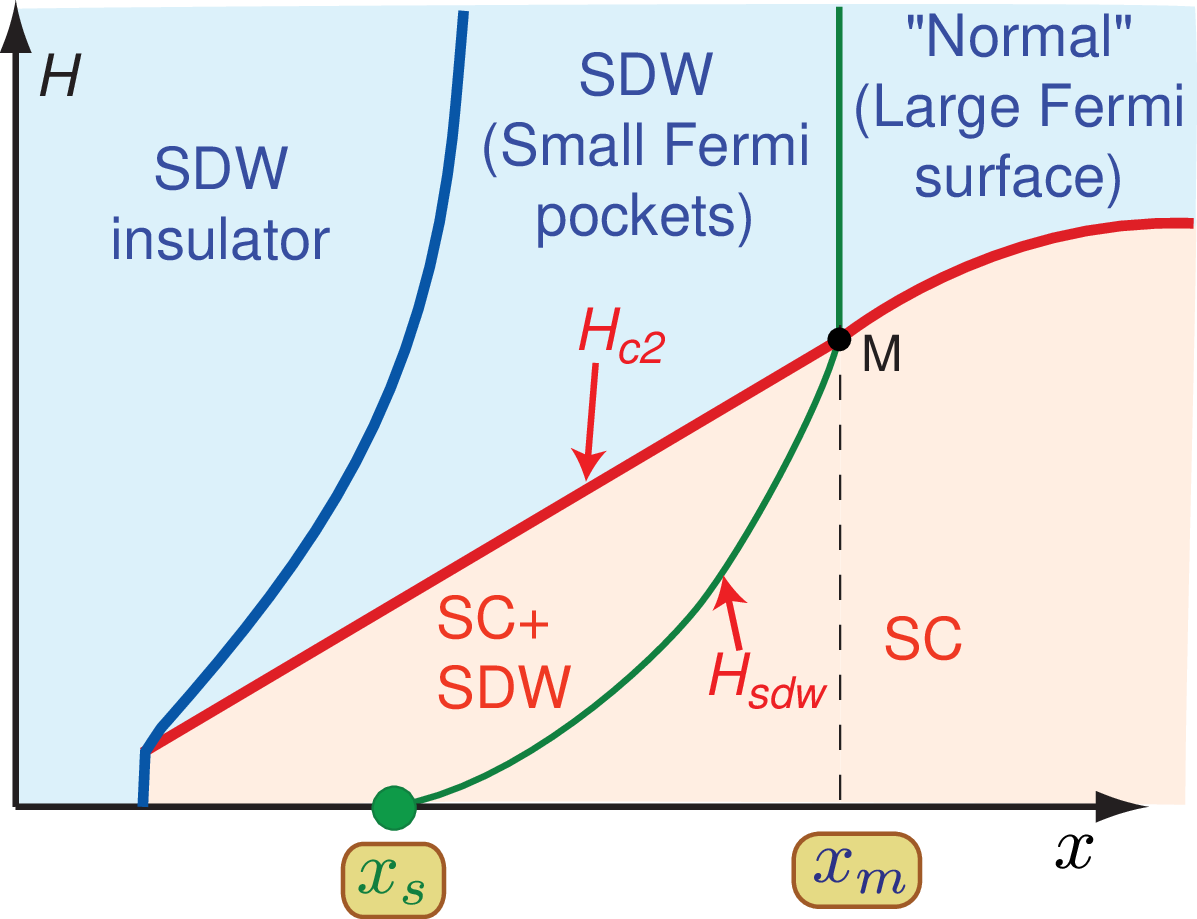}
\caption{A modified version of the $T=0$ phase diagram in 
Fig.~\ref{figdemler} showing the localization of the Fermi pockets across
a metal-insulator transition.}
\label{figloc}
\end{figure}
The structure of Fig.~\ref{figloc} is also consistent with earlier 
observations of Boebinger, Ando, Balakirev and collaborators \cite{greg}.
\item Andrey Chubukov (and also Louis Taillefer) pointed out that there should be an analog
of the line $H_{sdw}$ in Fig.~\ref{figdemler} at non-zero temperatures but $H=0$. We sketch such
a line, $T_{sdw}$ in Fig.~\ref{figtsdw}. 
\begin{figure}%
\includegraphics*[width=0.5\textwidth]{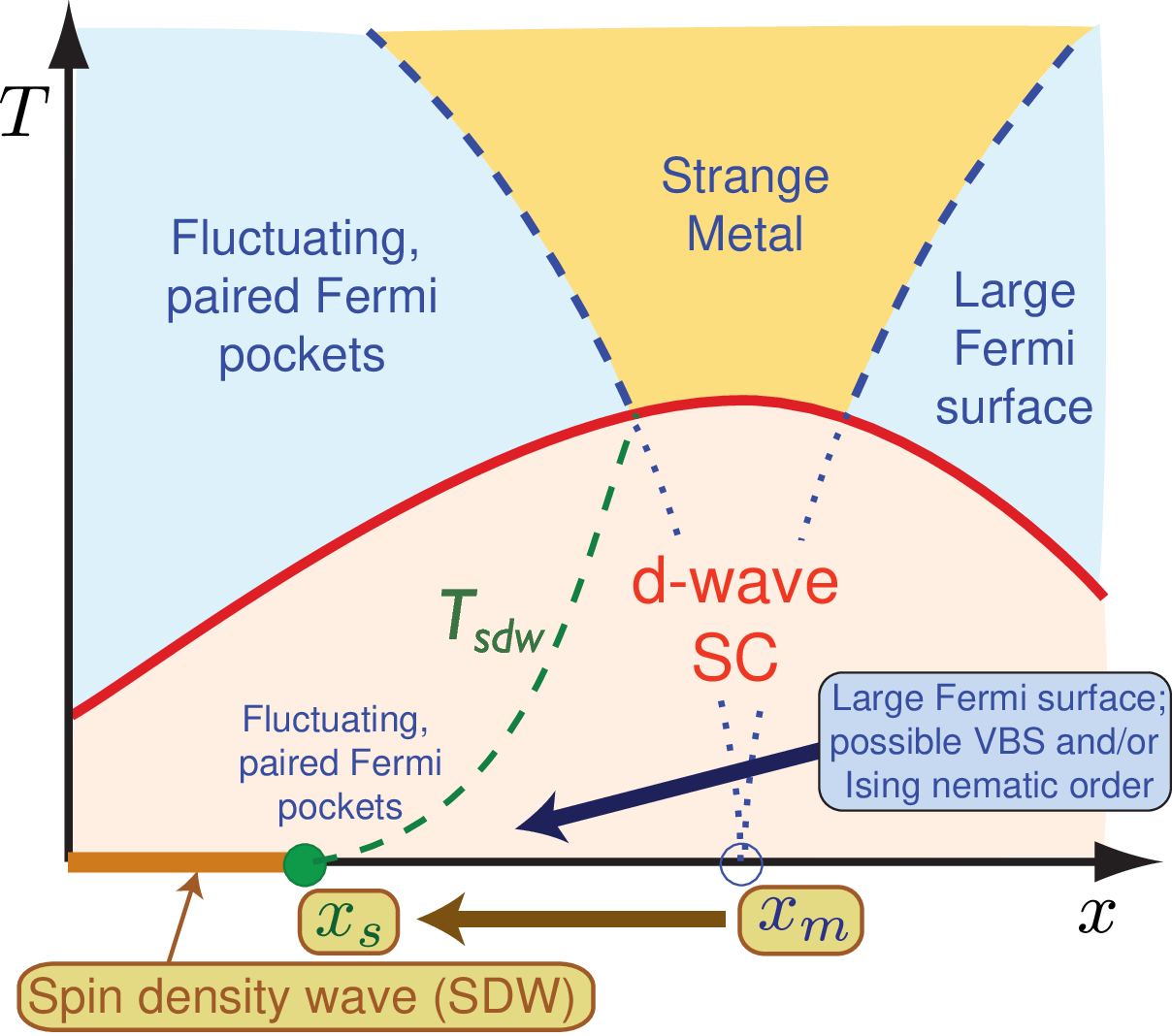}
\caption{Another view of the $H=0$ phase diagram in Fig.~\ref{figcross}. The crossover line $T_{sdw}$ is the analog of the
phase transition line $H_{sdw}$ in Figs.~\ref{figdemler} and~\ref{figtsdw}.}
\label{figtsdw}
\end{figure}
As we lower the temperature for $x_s < x < x_m$, the onset of superconductivity aborts the tendency towards SDW ordering
in the metal: this leads to a crossover at $T_{sdw}$ from Fermi pocket physics to large Fermi surface physics at the lowest energies.
Thus the nodal spectrum will be sensitive to Fermi pocket fluctuations above $T_{sdw}$ but not below it.
\item In all our phase diagrams, the SDW ordering is present as long-range order only at $T=0$.
This assumes the absence of appreciable inter-layer coupling: the SDW ordering has a large period, and differences
in the phase and direction of the ordering between adjacent layers will lead to an inter-layer coupling which averages to zero.
In contrast, for the two-sublattice N\'eel ordering present for $x < 0.02$ (not shown in our phase diagrams),  it is easier for 
the layers to lock-in, leading to three-dimensional ordering with a significant N\'eel temperature.
\end{itemize}

\begin{acknowledgement}
I am grateful to Andrey Chubukov, Eugene Demler, Victor Galitski, Ribhu Kaul, Yong-Baek Kim
Max Metlitski, Eun Gook Moon, Yang Qi, T.~Senthil, Cenke Xu, and Ying Zhang for collaborations
in which the ideas reviewed here emerged. I also thank Andrey Chubukov, Yoram Dagan, Gil Lonzarich, Christos Panagopoulos,
Suchitra Sebastian, and Louis Taillefer 
for numerous enlightening discussions;
Yoram Dagan drew my attention to the electron-doped case, and explained its connection
to the ideas presented here.
This research was supported by the NSF under grant DMR-0757145, by the FQXi
foundation, and by a MURI grant from AFOSR.
\end{acknowledgement}

%

\begin{thebibliography}{[99]}
\bibitem{spingap} In a $d$-wave superconductor, the Bogoliubov quasiparticles can create gapless spin excitations, but these
are away from the SDW ordering wavevector.
\end{thebibliography}
%

\end{document}